# Photoconductivity and photo-detection response of multiferroic bismuth iron oxide

Avneesh Anshul[a], Hitesh Borkar[b], Paritosh Singh, Prabir Pal, Sunil S Kushvaha, Ashok Kumar[a,*]

[a]CSIR-National Physical Laboratory, Dr. K. S. Krishnan Marg, New Delhi 110012, India

We report visible light detection with in-plane $BiFeO_3$ (BFO) thin films grown on pre-patterned inter-digital electrodes. In-plane configured BFO film displayed photocurrents with a 40:1 photo-to-dark-current ratio and improved photo-sensing ability for >15000 s (4 hrs) under small bias voltage (42V). Nearly sixty percent of the photo-induced charge carriers decay in 1.0 s and follow a double-exponential decay model. At 373 K the effect of light does not significantly increase the dark current, probably due to reduced mobility. Sub-bandgap weak monochromatic light (1 mw/cm$^2$) shows one fold increase in photo-charge carriers.

[*]Corresponding Author: Dr. Ashok Kumar (ashok553@nplindia.org); a and b author contributed equally to this work.

In the last decade multiferroic bismuth ferrite especially that doped at A- and B– sites has been an exciting material extensively investigated by researchers around the globe [1-3]. It different functional properties and cross coupled ferroic order parameters make it a potential candidate for nonvolatile memory elements, optical memory devices, above-bandgap photovoltaic devices, micro/nano electromechanical systems, magnetic field sensors, etc. [4-10]. The optical properties of BFO are also different from most of the perovskite ferroelectric due to its lower bandgap (2.6-2.9 eV) and conducting domain walls [11,12]. Clark and Robertson [13] predicted both direct and indirect bandgaps around 2.5 eV using band structure model, and afterward Ihlefeld et al. [14] reported a direct-gap transition at 2.74 eV, however recently we also found a direct bandgap of 2.55 eV at T = 295 K, an indirect band edge at 2.67 eV [5].

Conventional photoelectric (PE) devices are based on two basic principles: first, the generation of electrical charge carriers in the active medium under the illumination of light; and second, separation and detection of photoinduced charge carriers by in-built asymmetric potential of p-n junction or schottky junction or difference in work function between electrodes and semiconductor. The performance of the semiconductor PE devices is limited by the bandgap of the materials and the formation of narrow depletion regions or space charge regions at p-n junction. Recent studies revealed that limitations of semiconductor PE devices can be overcome in the ferroelectric semiconductor PE devices, which naturally possess built-in nano-scale p-n junctions of insulating domains and conducting domain walls [7,8,15]. Above bandgap open circuit voltages (OCV) have been shown in epitaxial BFO thin films utilizing the in-plane domains and domain walls [4]. Note that the basic mechanism of domain-wall PE effect is quite different than that of the bulk ferroelectrics (lack of inversion symmetry) [16]. We report the photon detection of solar spectrum and monochromatic visible light of BFO thin films having in-plane electrode configurations and modest illumination over wide range 1-100 mW/cm$^2$. A model of growth and decay of transient photocurrents is presented and discussed.

BFO thin films were grown on the pre-patterned inter-digital platinum electrodes of size 150 +/- 25 nm (height), 1900 μm (length), 20 μm inter-digital spacing, with 45 parallel electrodes each, procured from the NASA Glenn Research Center (electronic division) using pulsed laser deposition techniques. These capacitors are connected in parallel configurations. In-plane BFO film fabrication parameters and electrical measurement set-up are explained in our earlier reports. [5,17] A Precision Multiferroic tester (Radiant Technologies Inc.), and Keithley 236 source meter were used to measure the current-voltage for different measuring times and photocurrent stability over time. An Agilent B2900A source meter was used for the measurement of transient current, growth/decay current and photo-to-dark-current ratio as function of time and periodicity. Effect of weak monochromatic light on photoconductivity was measured with a two-probe station equipped with a TMc300 monochromator and a dual Xe/quartz halogen light source. Photoconductivity data were collected in dc mode using continuous illumination of different monochromatic light of desired wavelengths.

A schematic diagram of the pre-patterned inter-digital electrodes with an arrangement of bias electric field is shown in the inset of Fig.1. Photocurrent as a function of time is shown in Fig.1 over large acquisition times with a period of 250 s. It indicates robust reproducibility of switching states for a long time. Figure 2 (a,b) examine decay and growth of photo current for different switching times. We have used an in-plane inter-digital electrode configuration to measure the photocurrent growth and decay process. Sudden interruption and illumination of light allow the decay and growth of photo charge carriers over time. This is classical semiconductor characterization of photo-sensing elements and devices. BFO is known to be an n-type semiconductor at high electric field where a neutral oxygen vacancies can release one or two electrons in the conduction band [18], and a p-type semiconductor with introduction of Bi vacancies [19]. Here we established a model of photo-induced decay current that can be derived starting from the continuity equation in simplified form; $\frac{\partial I_c}{\partial t} = G - \frac{I_c - I_0}{\tau_m}$, the concentration of photo charge carriers is equal to the generation rate G minus the recombination rate, which is proportional to the photo-excited carriers ($\tau_m$ =life time of the carriers); I and t represent current and time [20, 21].

The solution of this equation gives us the decay in photocurrent using a double-exponential decay function: $I(t) = I_{fm} exp\left(-t/\tau_{fm}\right) + I_{sm} exp\left(-t/\tau_{sm}\right) + I_0$, where t is time, $\tau_{fm}$ and $\tau_{sm}$ are life times of the fast recombination and slow recombination carriers; $I_{fm}$ and $I_{sm}$, fast and slow recombination photo carriers current; $I_0$, offset current. The solution implies about 58% (+/- 3%) current ($I_{sm}$ fast recombination charge carriers) decay in one second, and remaining slow recombination charge carriers ($I_{sm}$) nearly reaches saturation (dark current) within 18 (+/- 3) seconds, depending on the saturation time. Fast and slow switching currents show similar responses within +/- 5%. The growth processes of photo charge carriers follow the equation, $I(t) = I_{fm} exp\left(1 - {-t}/{\tau_{fm}}\right) + I_{sm} exp\left(1 - {-t}/{\tau_{sm}}\right) + I_0$. The growth and decay process of the experimental data and a good fit from this model are given in the Figure 2 (c). The photocurrent growth process indicates that 20% of fast photocurrent reaches its maximum value within 1.0 s; however, remaining 80% from minority slow carriers reaches to maximum value within 24 (+/-5) seconds. The growth of photocurrent is slow compared to its decay.

Figure 3 (a,b) show current versus voltage (I-V) of an in-plane BFO thin film with and without light, using custom-designed pulse voltage system (pulse width ~ 1 ms, soak time ~ 100 ms, step delay time ~ 100 ms, & different measure times (10-10000 ms)) at 298 K and 373 K, respectively. Measure time is the time over which to measure the current while the sample is at the step voltage. I-V data were collected for different measure times to check the current response at low applied voltages (below 25 V). It can be seen from the Fig. 3(a) that fast measure times do not provide smooth currents, however slow measure times show smooth behavior of currents. I-V graphs show a similar trend of current growth across both electrodes. The important observations are as follows: 1-2 orders of change in currents were observed under visible light over wide range of field and effect of light was almost negligible at higher T (near 373 K). In the literature several conduction mechanisms have been proposed to find out for conduction in the dielectric: Schottky emission (SE), Poole-Frenkel emission (PF), and space charge limited conduction (SCLC) [22-24]. These three basic mechanisms represent the interface, bulk, and free-charge carriers injection as function of applied external fields and temperatures, respectively. In case of SE, log ($I/T^2$) vs $V^{1/2}$ should gives a straight line having slop near to the optical dielectric constant of the materials [25].

In present case Fig 4(a), at low field and temperature, BFO follows non-linear current-voltage behavior, ruled out the SE effect; it fits linearly at high temperatures and electric fields, but the slope of curve (~ 0.50 +/- 0.05) does not match the optical dielectric constant of BFO, and more importantly rules out SE as the dielectric constants cannot be less than one. Fig 4 (b) represents the PF formalism fitting of current-voltage data; almost similar trends like SE can be seen. At high temperatures and electric fields, log (I/V) vs. $V^{1/2}$ fits linearly with slope of curve (~ 0.30 +/- 0.05) far away from optical dielectric constant of BFO also ruled out possible PE mechanism.

These experimental results may be explained on the origin of randomly oriented domains (bulk) and domain walls (boundaries) for polycrystalline ferroelectric systems. [21,26,27]. These randomly oriented domains and domain-walls guide in-plane photocurrent. Physical model based on complex domains and domain walls of epitaxial BFO thin films are presented elsewhere [21,28]. In polycrystalline ferroelectric films/ceramics, photo current has been characterized to series connections of different grains/grain boundaries or domains/domain walls [29]. The conducting domain walls become more semiconducting or insulating with increase in temperature which may reduce the charge generation near the interface. Figure 5 shows the temperature dependent behavior of dark and light current for two different stress voltages (42 and 98 V). The ratio of light-to dark-current decreases with increase in temperature and becomes almost (1:1) near 373 K. The intrinsic carrier concentration increases the dark saturation (recombination) current with increase in temperature [20]. The photocurrent charge carriers such as electrons and holes, are the commonly classified as drift and diffusion current. Temperature dependent drift and diffusion current enhanced phonon scattering which in turns destroy the photo-charge carriers. Space charges, oxygen vacancies and trapped charges may also be responsible for the fast recombination at elevated temperature [30].

To guarantee device lifetime, one has to understand the charge/current retention and reproducibility of photocurrent under illumination light. Inset of Fig. 5 shows dark and light current under an electric field stress for long time (> 15000 s) across the photo-current collection terminal. It indicates an increase in current of at least 1-2 orders of magnitude under weak light. Interestingly the switching ratio improves under the long stress time. Under small stress voltage (42 V) the dark current slowly decreases with time (which is rather good for device applications) and finishes with a high switching ratio.

Figure 6 shows the spectral response of BFO as function of wavelengths, it indicates sharp increase of photocurrent at 370 nm (~ 3.35 eV) far above the bandgap which energetically persist until the bandgap (~ 2.63 eV) [5]. A small kink near the 650 nm (1.91 eV) was observed due to the presence of oxygen vacancies. The presence of oxygen vacancies is common in most of the oxides fabricated at elevated temperature. Similar effect was observed in the photoluminance spectra of the BFO thin films [5]. Effect of different intensity of the Xe/quartz halogen light on the photocurrent has shown in the inset of figure 6 that suggests photocurrent increases monotonically with increase in light intensity. Photoconductivity data were obtained in dc mode using continuous illumination of different weak monochromatic light (1 mw/cm$^2$) of desired wavelengths as can be seen in Figure 7 (a-c). Photocurrents for sub-bandgap monochromatic visible wavelengths is at least one order higher compare to the dark current, the magnitude of photocurrent increases until 400 nm wavelength, matched well with the spectral response data. The ratio of photo-to-dark-current is 6:1 for 400 nm wavelength, 5:1 for 470 nm wavelength and 40:1 for Xe/quartz halogen light, note that the intensity of Xe/quartz halogen light is 100 time higher than the monochromatic light ranging from 400 to 550 nm wavelengths. Seidel et al. explained BFO in-plane photocurrent with similar magnitude on the basis of a series connection of textured domains and domain walls [28]. Interestingly, sharp rise and decay of photocurrents were observed for 400 nm monochromatic wavelength (Fig. 7 (c)) contrary to the use of Xe/quartz halogen (Fig.2), this may be due to the presence of different energy photons in the visible wavelengths and hence different recombination process. The time constants for growth and decay of the photocurrents to reach the saturation current using Xe/quartz halogen source were 25 (+/-2) seconds and 19 (+/-2) seconds, respectively. However, small time constants, 6 (+/- 2) seconds (for growth) and 8 (+/-2) seconds (for decay), of the photocurrents were observed using monochromatic light to reach the saturation current.

In summary, the skin layer of the BFO is reasonable sensitive to detect the visible light at ambient atmospheric conditions. In-plane photo current is 50-100 x higher than the dark current. The photo-current exponential growth and decay with time fit well to double exponential fast and slow recombination photocurrent model; however, the photo charge carriers decay very quickly ( ~ 60% @ 1.0 s) in comparison to their growth (~ 20% in 1.0 s). Spectral response demonstrates high intensity photocurrent in the Sub-bandgap region. Photocurrent shows sharp rise and decay with moderate photo-to-dark-current ratio under the illumination of weak monochromatic light (1 mw/cm$^2$) above the bandgap.

**Acknowledgement:**

Hitesh Borkar would like to acknowledge the University grant commission UGC-CSIR (UGC-JRF) to provide fellowship to carry out Ph. D program. Experimental support by Prof. R S Katiyar (UPR) is greatly acknowledged.

**Figure Captions**

Fig.1(a) shows the photocurrent as a function of time under the illumination of weak visible light at an interval of 250 s. Schematic outlook of in-plane configuration is given in the inset.

Fig. 2 (a,b) show decay and growth of photo-current for 100s, 250s and 500s of saturation time, respectively. Figure 2(c) shows the double exponential fitting of growth and decay current for 100 s of switching time interval.

Fig. 3 Current voltage response of in-plane BFO thin film for different measure times with and without illumination of light (a) at 298 K, (b) at 373 K. Slow measure times give better current trend as function of applied voltage below 25 V at 298 K.

Fig. 4 (a) ln ($I/T^2$) vs $V^{1/2}$ fitting of current and voltage data (SE fitting) and (b) ln (I/V) vs $V^{1/2}$ fitting of current and voltage data ( PF fitting) in dark and light at various temperatures ranging from 298 K-373 K.

Fig. 5 shows the effect of light on the charge carriers as a function of temperature with different stress voltages (a) 42 V and (b) 98 V. Photocurrent as a function of cumulative time at different stress voltage (inset).

Fig.6 shows spectral response as a function of wavelengths, photocurrent under different power of visible light (inset).

Fig.7 (a) Photocurrents under the illumination of different monochromatic wavelengths at 42 V dc bias, (b) Performance of photo-to-dark-current ratio under different monochromatic light, (c) sharp rise and fall of photocurrents for 400 nm wavelength.

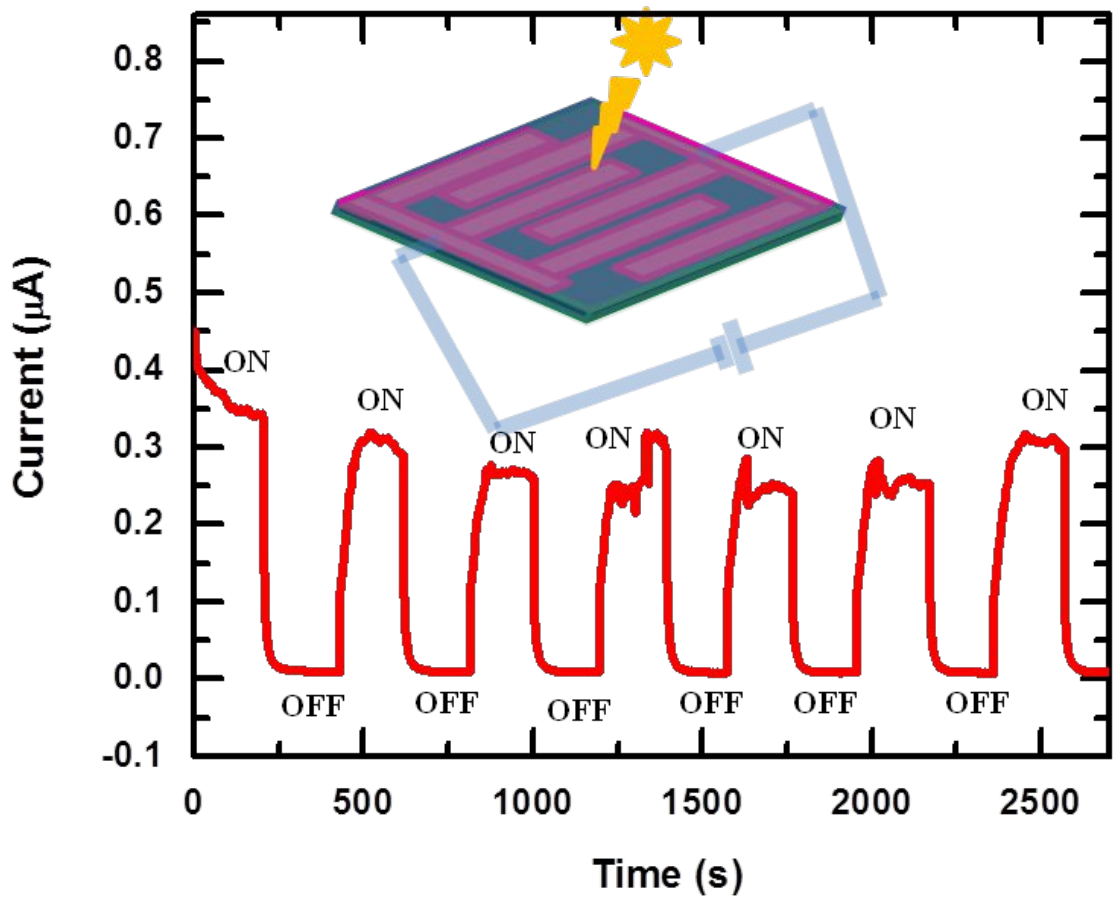

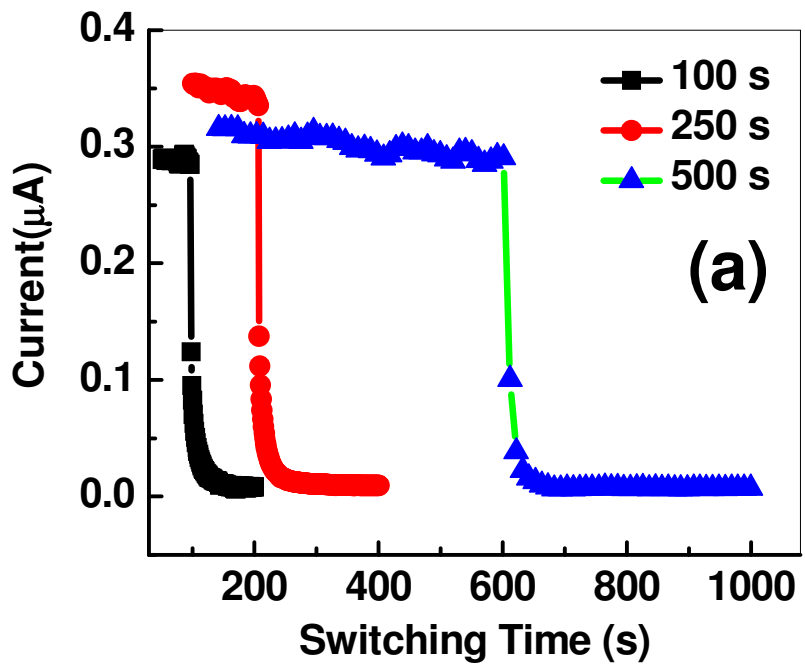

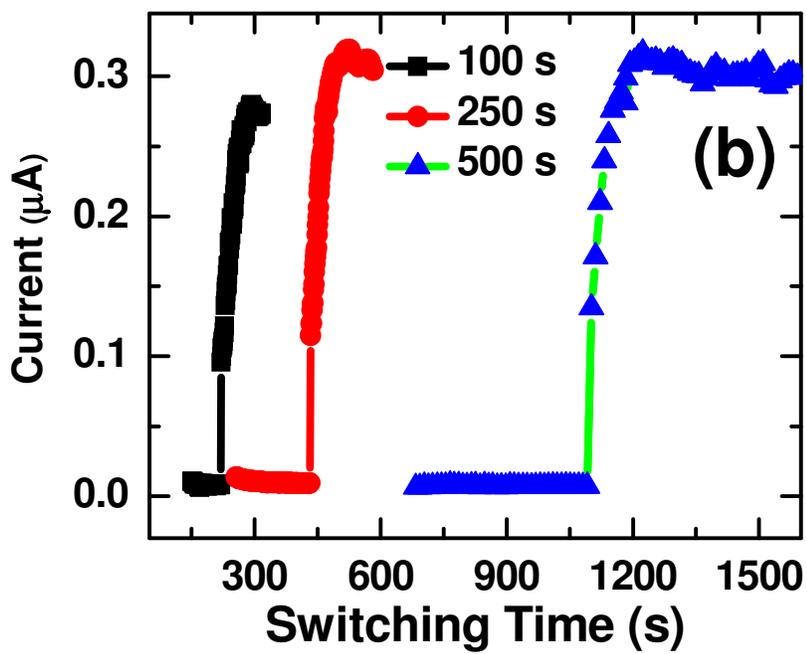

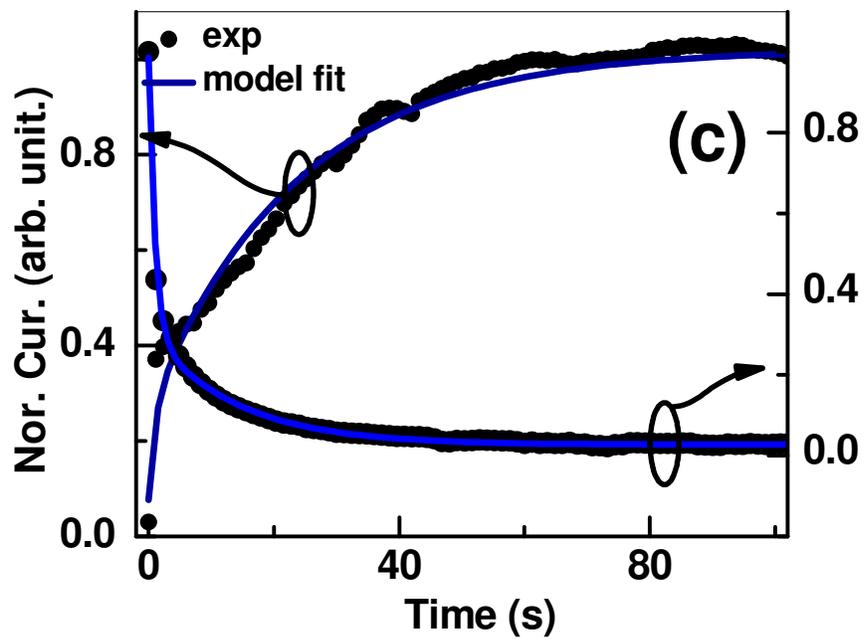

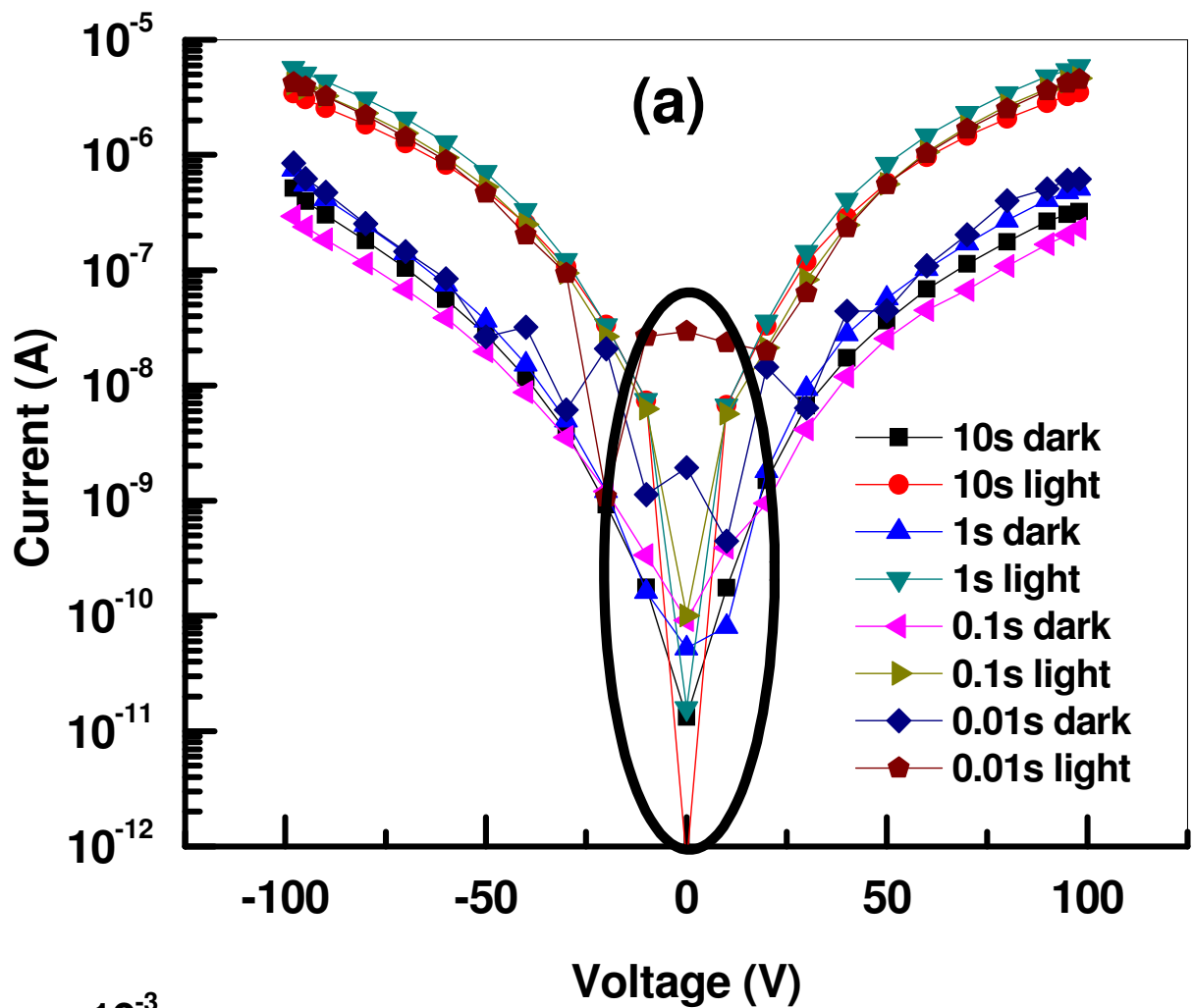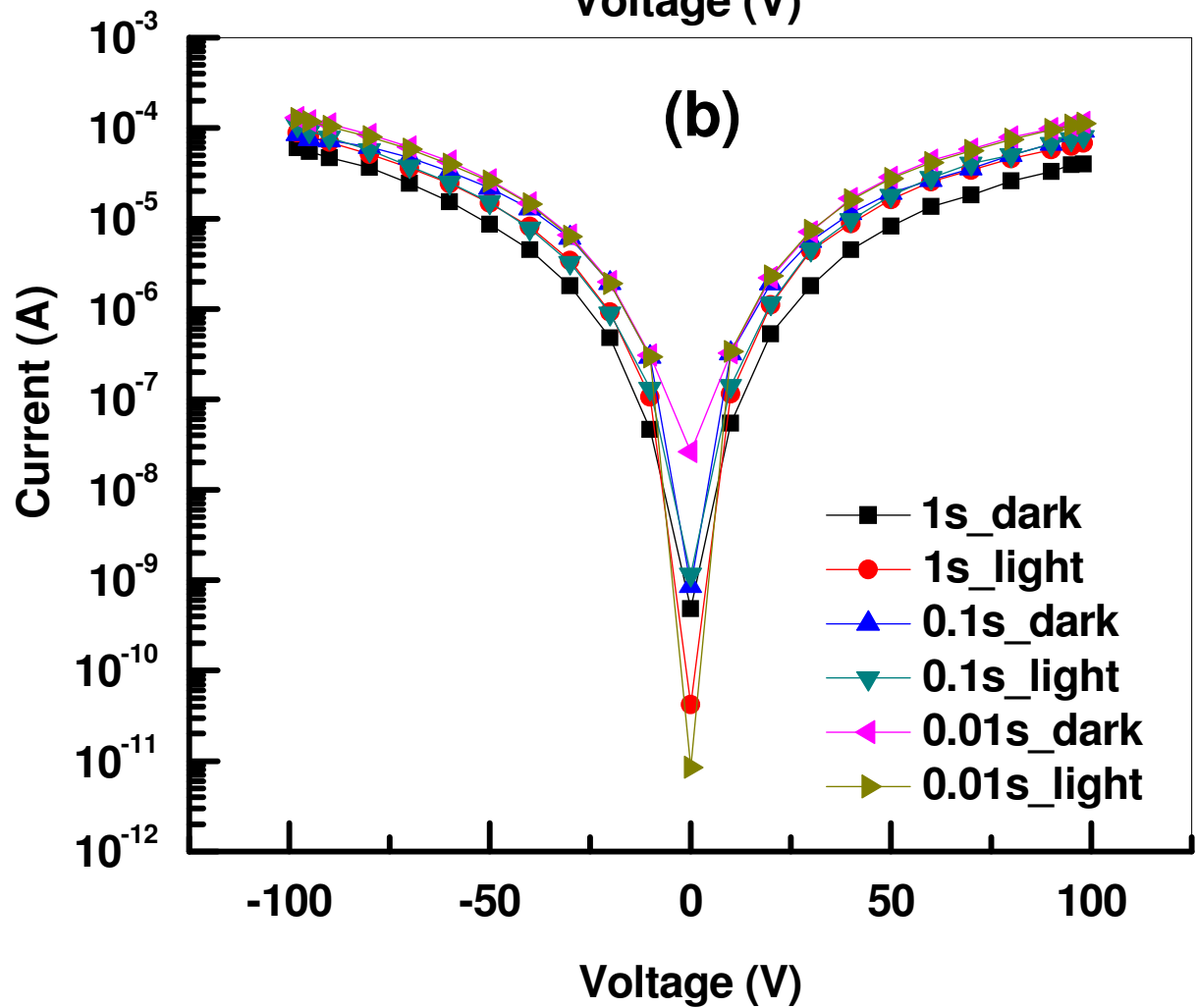

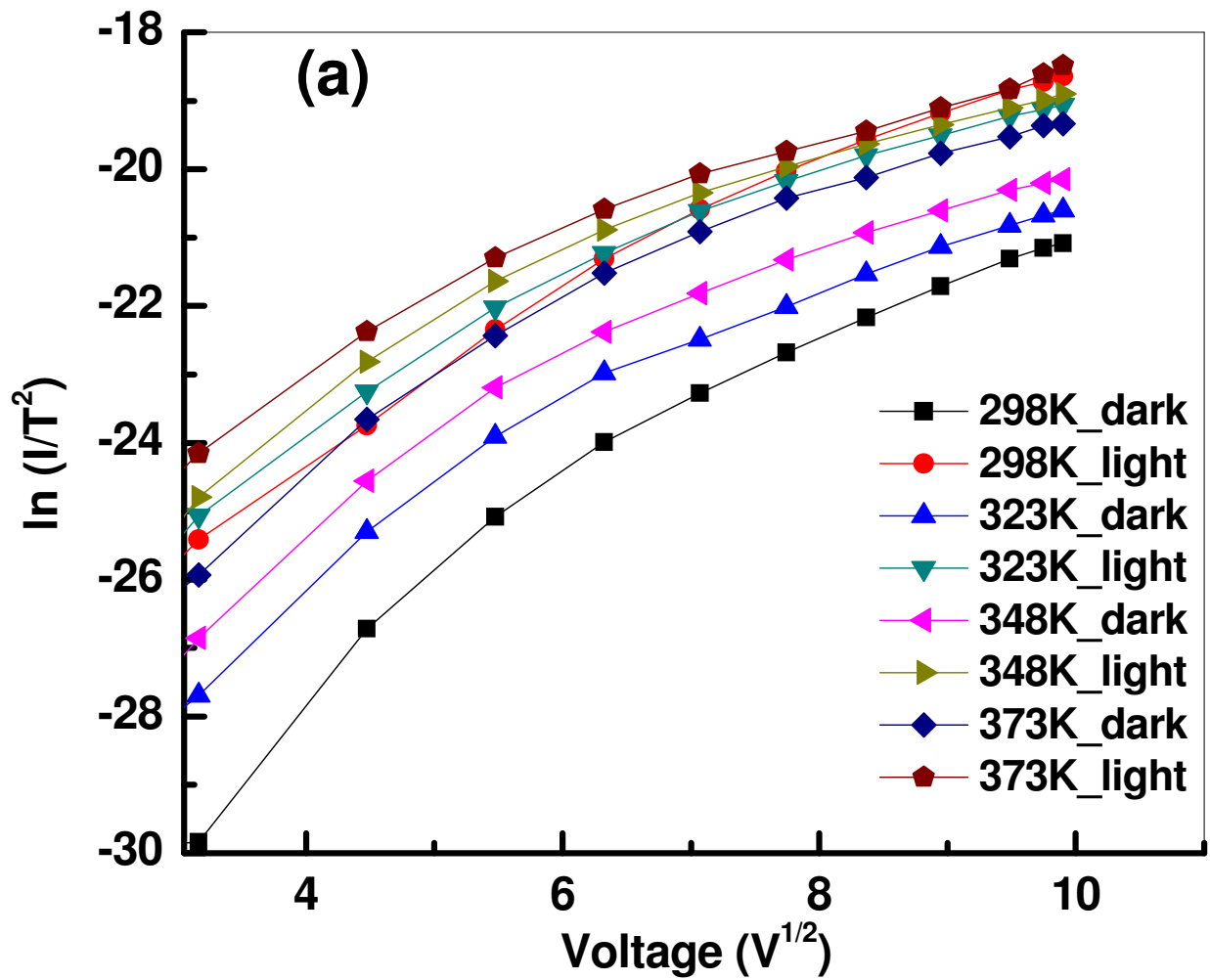
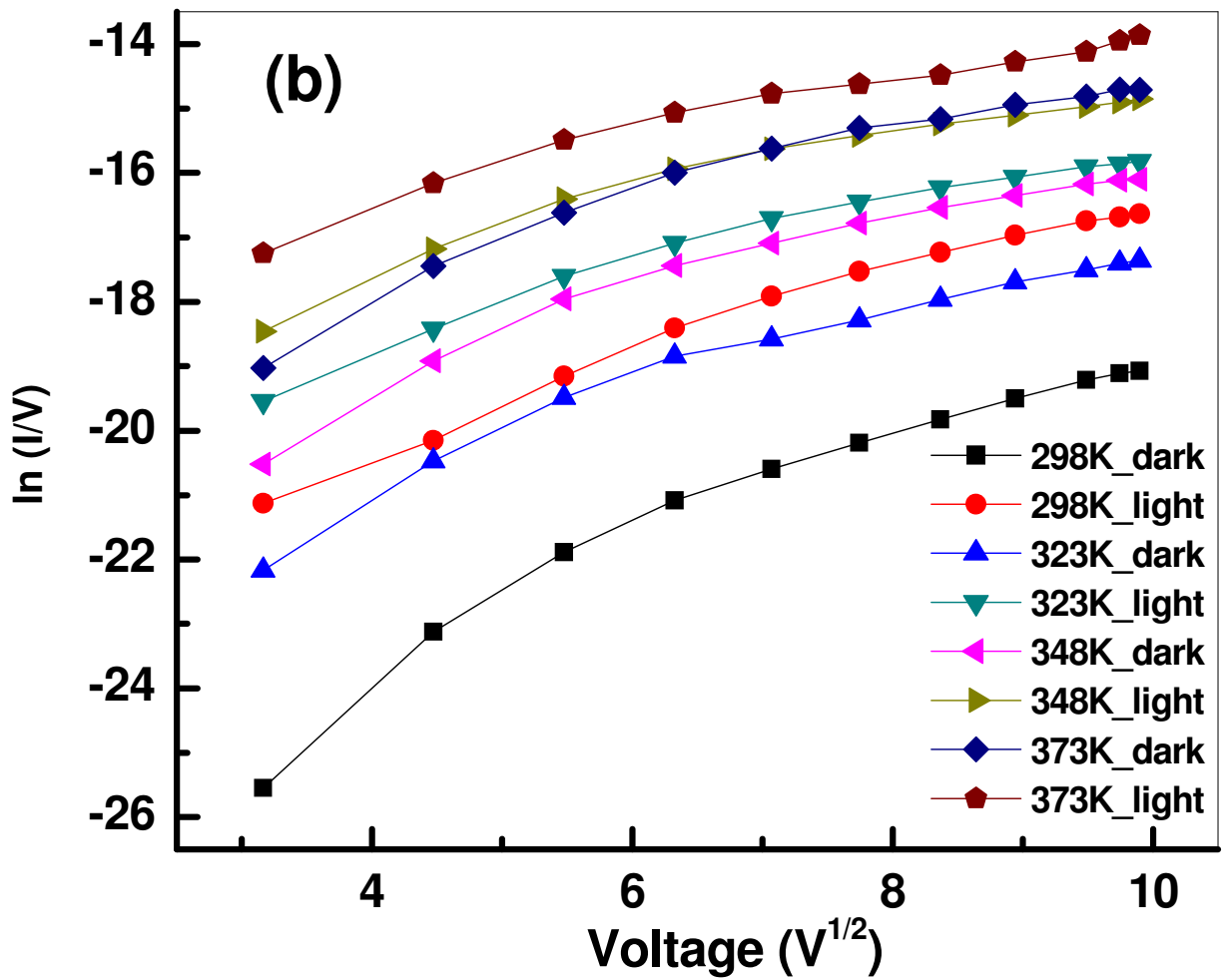

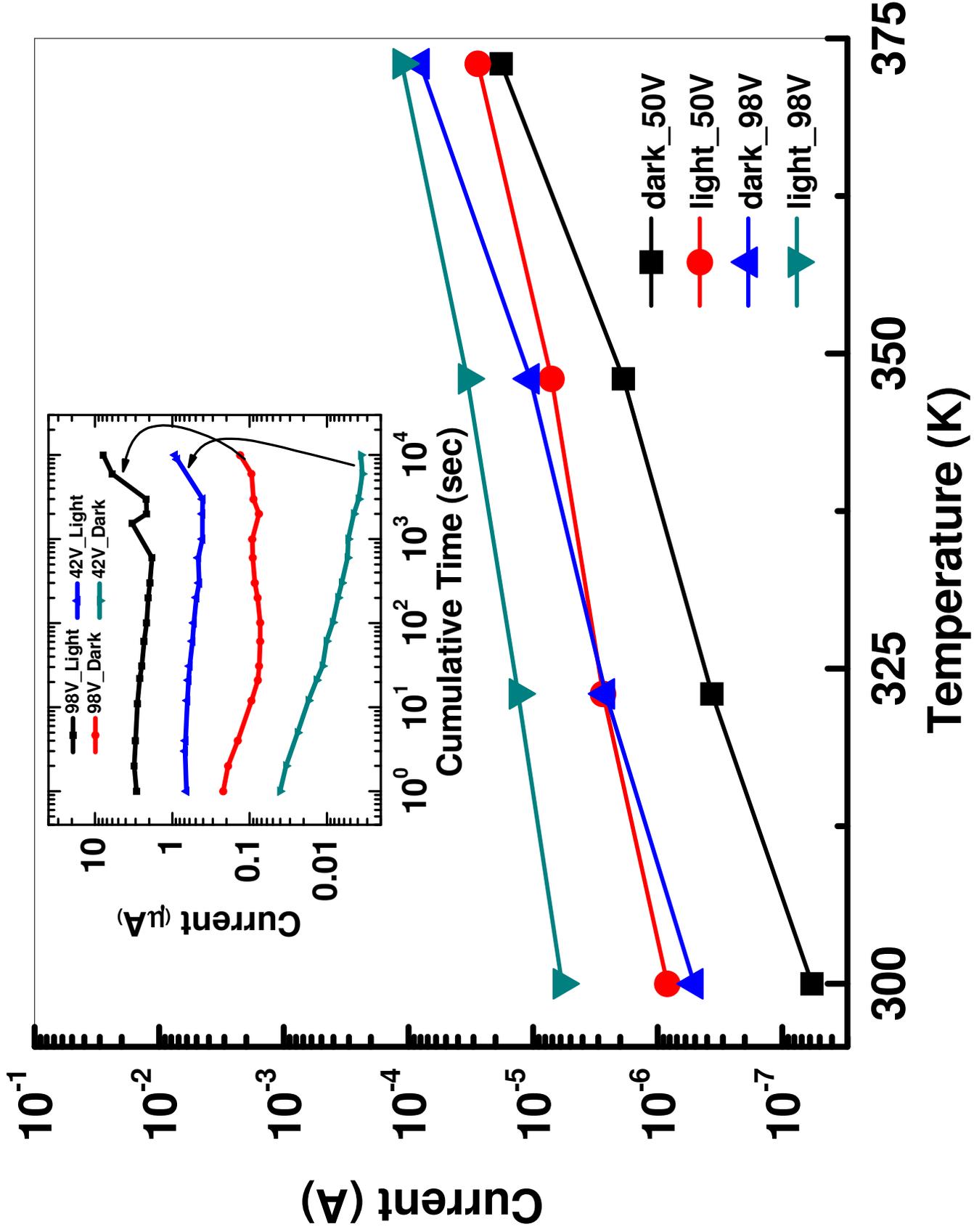

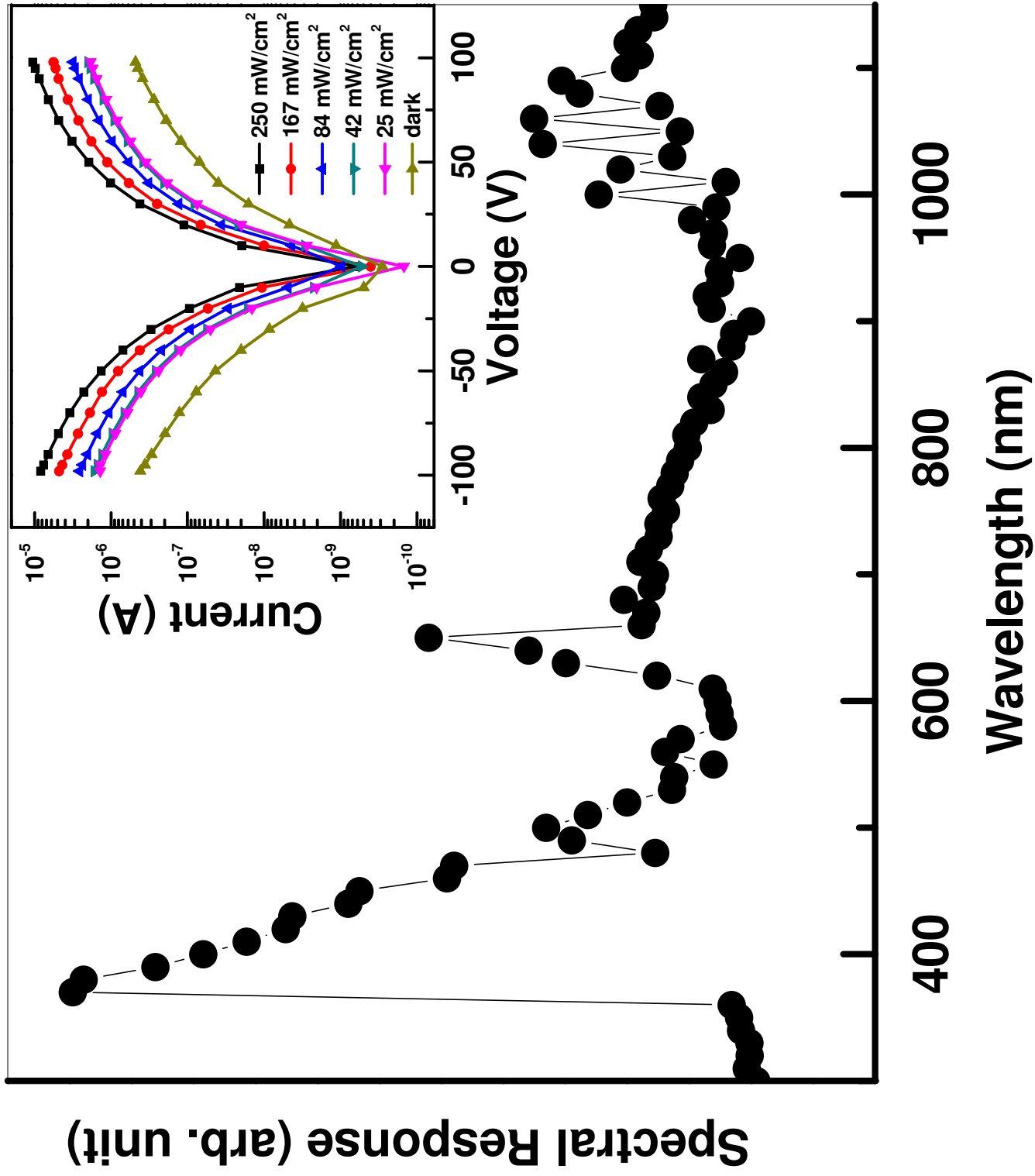

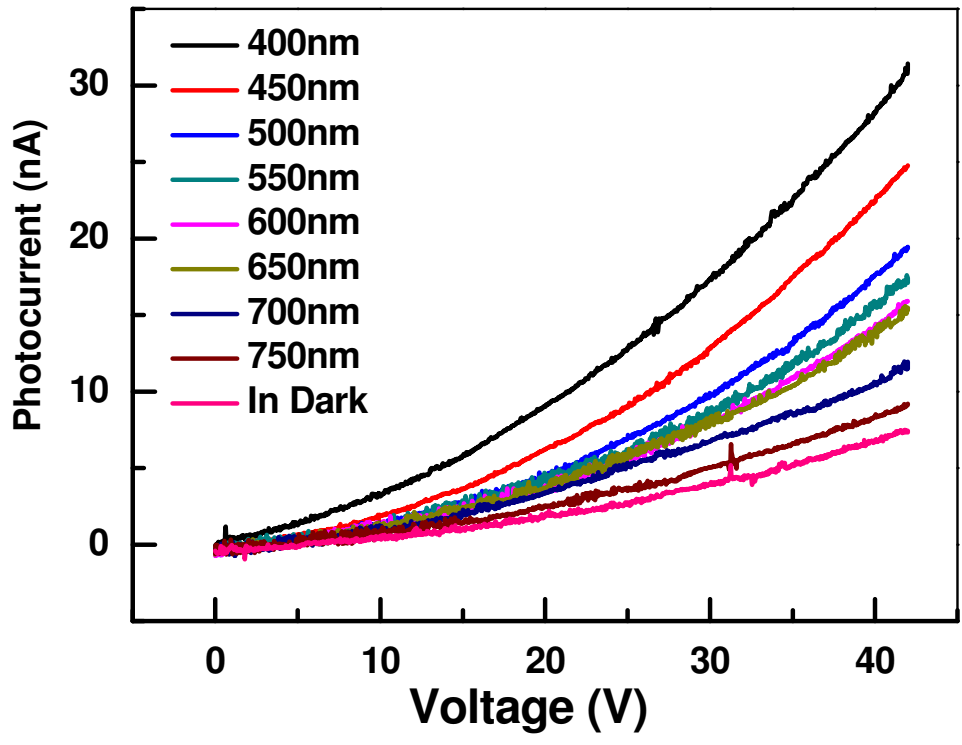
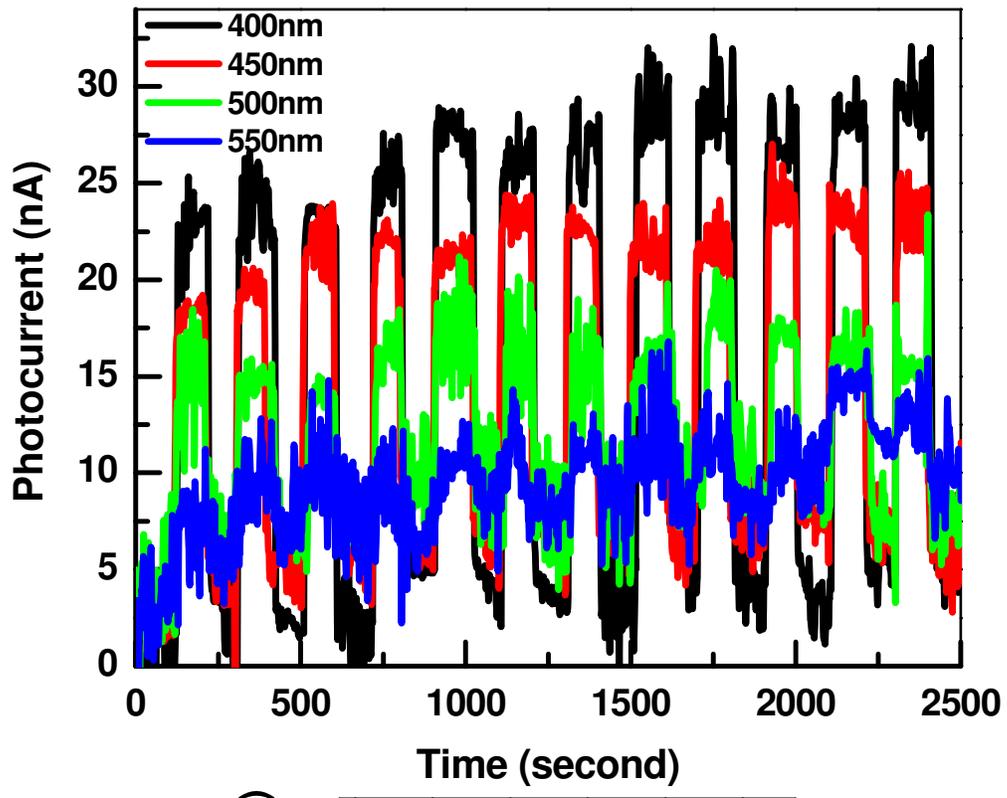
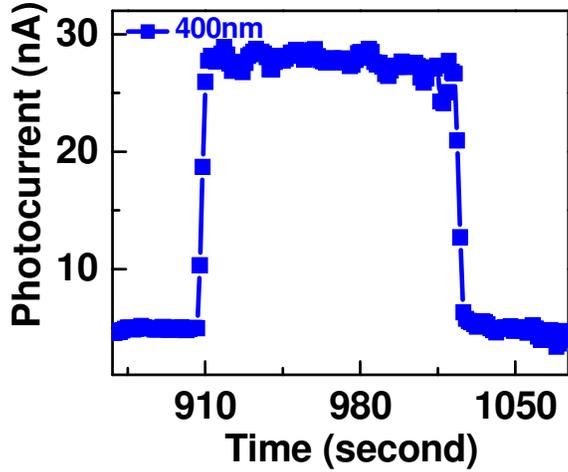